# Research on Multi-Objective Planning of Electric Vehicle Charging Stations Considering the Condition of Urban Traffic Network

*Limeng Wang [a], Chao Yang [b,\*], Yi Zhang [a], Fanjin Bu [a]*

[a] *Key Laboratory of Modern Power System Simulation and Control & Renewable Energy Technology, Ministry of Education (Northeast Electric Power University), Jilin 132012, China*

[b] *State Grid Harbin Power Supply Company Substation Secondary Maintenance Center, Harbin 150036, China*

**Abstract:** As an important supporting facility for electric vehicles, the reasonable planning and layout of charging stations are of great significance to the development of electric vehicles. However, the planning and layout of charging stations is affected by various complex factors such as policy economy, charging demand, user charging comfort, and road traffic conditions. How to weigh various factors to construct a reasonable model of charging station location and capacity has become a major difficulty in the field of electric vehicle charging facility planning. Firstly, this paper constructs the location and capacity optimization model of the charging station with the goal of maximizing the revenue of operators and minimizing the user's charging additional cost. At the same time, the road time-consuming index is introduced to quantify the impact of road congestion on the user's charging additional cost, so as to effectively improve the user's satisfaction during charging. Then, aiming at the charging station planning model, a non-dominated sorting genetic algorithm with an elite strategy (NSGA-II) based on chaos initialization and arithmetic crossover operator is proposed. Finally, taking the Haidian District of Beijing as the simulation object, the results show that compared with the situation of urban traffic networks not considered, the model proposed in this paper significantly reduces the cost of lost time of users by 11.4% and the total additional cost of users' charging by 7.6%. It not only ensures the economy of the system, but also effectively improves the charging satisfaction of users, which further verifies the feasibility and effectiveness of the model, and can provide a reference for the planning and layout of charging stations in the future.

## 1 Introduction

Under the background of China's "carbon peak and carbon neutrality" strategy, electric vehicles have gradually become an indispensable part of modern transportation due to the advantages of cleanliness, efficiency, and no pollution [1-3]. As an important supporting facility for electric vehicles, the reasonable planning and layout of charging stations will help to further expand the scale of the electric vehicle market [4,5]. However, there is still a certain gap between the construction scale of charging stations and the ownership of electric vehicles. The charging difficulty of users is still an important factor restricting the development of electric vehicles. According to data from the China Electric Vehicle Charging Infrastructure Promotion Alliance, as of September 2021, the number of new energy vehicles in China has reached 6.78 million, while the number of charging infrastructures is only 2.223 million, and the ratio between electric vehicles and charging piles is about 3.05:1[6]. This is still far from the strategic goal of "the 1:1 ratio between electric vehicles and charging piles" proposed by the National Energy Administration, and the planning task of charging stations is still arduous.

The location and capacity modeling of the charging station is affected by multiple factors. It is not only necessary to consider the investment cost, operation cost and other cost factors of the charging station, but also pay attention to the satisfaction of users during charging, so as to weigh multiple factors to select the best layout scheme of the charging station. At present, experts and scholars around the world have carried out a series of researches on the planning of charging stations and achieved certain results [7,8]. Reference [9] analyzes the cost components of the income of the operator and proposes an upper-layer model aiming at the maximum profit of the operator and a lower-layer model aiming at the maximum user satisfaction index, which provides a new idea for the solution of the planning problem for the charging station. Reference [10] constructs a total social cost model including comprehensive economic cost and environmental cost and calculates the total operating cost of charging stations under various distribution conditions. The case results show that the total cost is very sensitive to the number of charging stations and the charging possibility of electric vehicles every day. Based on the research on the traditional location of charging stations for the electric vehicle, reference [11] introduces service risk factors, including service capacity risk and user anxiety risk, and establishes a location model of charging station for the electric vehicle considering service risk. Considering the charging convenience of users, reference [12] proposed a two-stage planning method of charging stations for the electric vehicle combining charging stations and decentralized charging piles, so as to balance the economy of operators and the convenience of users. Reference [13] comprehensively considers the interests of both the charging station operator and the electric vehicle user and establishes a charging station location and capacity model with the minimum cost as the optimization objective. In combination with the road traffic flow information in the planning area, the node weight is set according to the size of the traffic flow to determine the service scope of the charging infrastructure. In addition, in order to more clearly describe the user's recognition of charging at the charging station, some researchers have proposed the concept of user charging satisfaction. According to the different charging times of users in the charging station, reference [14] uses the improved sigmoid function to define the satisfaction of electric vehicle users with the time spent charging in the charging station and establishes a two-level programming model for a centralized charging station that considers the investment cost of operators and the satisfaction of users. The above research not only considers the total operating cost of the charging station but also pays more attention to the satisfaction of users when charging, which has achieved good planning results.

Compared with the overall planning of the charging station, the capacity configuration in the electric vehicle charging station is also of great significance to the economic operation of the system and the cost saving for users [15-17]. An excellent charging station capacity configuration scheme can not only save the initial investment cost of the charging station operator and help reasonably allocate the existing charging resources, but also effectively reduce the charging time and queuing time of users, and minimize the probability of station congestion. Reference [18] proposes a Baskett, Chandy, Muntz and Palacios (BCMP) queuing network model, which regards each charging station as a service center with multiple charging piles, while electric vehicles are modeled as customers of the service center. Based on the stable distribution of the number of electric vehicles in each charging station, the statistical data on the charging demand of electric vehicles can be obtained, determine an effective charging station capacity configuration scheme. Reference [19] considers the charging power of electric vehicles and the SOC at the time of arrival/departure when estimating the charging time of users, and models the charging station as a multi-level M/G/s queuing network, wherein the waiting time of users is calculated by considering the charging requirements of different electric vehicle models in the Spanish market. The case results show that the charging station configuration scheme proposed in this paper can effectively improve the working efficiency of the charging station, thereby achieving the shortest waiting time and the maximum utilization rate of the charging station. Meysam Hosseini established a charging station planning model by considering capacity, charging time, and waiting time [20]. Based on the expected arrival rate and expected charging time, reference [21] established an M/M/n/K queuing model considering parking space constraints to optimize the capacity of the charging station under the condition of ensuring the charging probability loss of electric vehicles. When the capacity configuration of the charging station is carried out in the above documents, the economy of the operation of the charging station is taken into account, which effectively reduces the total cost of the charging station.

How to select the appropriate method to solve the model under the corresponding constraints such as determining the charging demand of electric vehicles, road network, and charging radius, to achieve the optimal planning goal has become the focus in the field of electric vehicles. With the gradual deepening of research, intelligent optimization algorithms have gradually been widely used in power system optimization problem like electric vehicle charging station planning, such as the NSGA-II algorithm [22], multi-objective particle swarm algorithm (MPSO) [23-26], and multi-objective ant-lion optimization algorithm (MOALO) [27-29], etc. NSGA-II algorithm has high performance in finding a set of Pareto solutions, so it has gradually become one of the most commonly used multi-objective optimization algorithms in different application scenarios. Since the planning model of charging stations is a complex nonlinear multi-objective optimization problem, the NSGA-II algorithm has good convergence speed and robustness in solving such problems, so it is favored by experts and scholars. Reference [30] first used an efficient NSGA-II algorithm to find the location and capacity of the fast charging station and distributed power generation, to optimize the loss of electric vehicle users, network power loss, and development cost of charging stations for the electric vehicle and improve the voltage distribution of power distribution system. Reference [31] proposed an optimization planning model of charging stations that comprehensively considered the impact of public service infrastructure, charging station construction cost, distribution network integration cost, and distributed generation. At the same time, the NSGA-II algorithm was used to solve the model to obtain the optimal Pareto frontier. Reference [32] established the planning model of charging stations for the electric vehicle considering the strongest charging service capability and the smallest distribution network loss and used the traditional NSGA-II algorithm to solve it. Finally, the superiority of the model and algorithm is effectively verified by a simulation case.

However, although most of the studies have considered the additional cost caused by the user's charging satisfaction, they have not analyzed the impact of traffic congestion in the planned area on the user's charging behavior. This leads to a deviation between the calculation of the additional cost of charging when the user travels to the target charging station and the actual situation. At the same time, when optimizing the capacity of the charging station, most of the optimization schemes are put forward for the whole charging station, without considering the charging characteristics of the charging piles inside the charging station. The suitable installation scenarios of charging piles of different powers are different, so it is necessary to combine charging piles of different powers in the charging station. In addition, in terms of the solution algorithm, although the NSGA-II algorithm has strong search ability and the Pareto solution set obtained is close to the optimal, it also has some shortcomings, such as low computational efficiency and poor distribution uniformity of the solution set. Therefore, the NSGA-II algorithm should be further improved to help solve the optimal planning scheme of charging stations. Based on summarizing the existing research results at home and abroad, this paper conducts in-depth research and analysis on the location and capacity problem of charging stations. The main innovations of this paper are summarized as follows:

1) The impact of road traffic conditions on the user's charging additional cost is analyzed, and the numerical relationship between the actual road conditions in the planning area and the user's charging additional cost is characterized by defining a road time-consuming index, which helps to further improve the economy of the scheme from the user side, and also effectively guarantees the rationality and applicability of the scheme.

2) The capacity configuration of the charging station is creatively transformed into the specification and capacity configuration of the charging piles in the station, and the installation quantity of different power charging piles is set according to the charging demand in the actual area. The above method can not only save the cost of the charging station but also better meet the actual situation of the spatial distribution of charging demand in different regions, thus further improving the travel satisfaction of users.

3) Aiming at the constructed location and capacity optimization model of charging stations, an improved NSGA-II algorithm based on chaotic initialization and arithmetic crossover operator is proposed, which can significantly improve the global convergence speed of the NSGA-II algorithm and also realize the diversity of population differences, making the obtained Pareto solution set closer to the optimal solution.

## 2 The location and capacity planning model of the charging station

The location and capacity plan of charging stations for the electric vehicle is not only related to the operator's income but also directly affects the satisfaction of users when charging. However, these two goals are contradictory and mutually restrictive. On the one hand, the charging station operators reserve the number of charging stations to reduce the cost of charging stations, which affects the charging satisfaction of users to a certain extent. On the other hand, the number and capacity of charging stations in the planned area need to be greatly increased to minimize the user's charging additional cost and effectively improve the charging satisfaction of users, but it will lead to a significant increase in the cost of operators. Therefore, it is necessary to comprehensively weigh the two to determine the best location and capacity plan of charging stations.

### 2.1 Objective function

In this paper, the optimization goal is to maximize the revenue of operators and minimize the user's charging additional cost, and a location and capacity planning model of charging station is established to maximize the charging satisfaction level of users while ensuring the income of operators. Therefore, the objective function is established as follows:

### 2.1.1 The revenue of the charging station operator

The revenue of the charging station operator is mainly composed of the total construction cost of charging station and the total operating income of the charging station. The calculation formulas are as follows:

$$\max f_1 = I - C_1 \tag{1}$$

$$I = T_{year} \sum_{i=1}^{D} \alpha NUM_i cap \left( SOC_{ref} - SOC_0 \right)\left( c_s - c_p \right) \tag{2}$$

where $I$ is the annual operating income of the charging station operator, $C_1$ is the total annual operating cost of the charging station operator, $T_{year}$ is the average annual operating days of the charging station, $D$ is the set of all roads in the planning area and it is a constant, $\alpha$ is the proportion of electric vehicles that need to be charged every day in the planning area, $NUM_i$ is the total number of electric vehicles on the $i$-th road, $cap$ is the average rated capacity of the electric vehicle, $SOC_{ref}$ is the state of charge threshold of the electric vehicle during the charging process, $SOC_0$ is the initial value of $SOC$ when the electric vehicle is ready to be charged, $c_s$ is the charging electricity price of the user, and $c_p$ is the electricity purchase price of the charging station.

The annualized total cost of the charging stations operator is mainly composed of the construction cost of the charging station, the purchase cost of the charging pile, and the operation and maintenance cost of the charging station. The calculation formula is shown in formula (3):

$$C_1 = \sum_{j=1}^{N} \left[ \frac{r(1+r)^T}{(1+r)^T - 1}(C_j + C_{cp,j}) + C_{om,j} \right] \tag{3}$$

where $N$ is the total number of charging stations to be built in the planning area, $r$ is the discount rate of the charging station, $T$ is the maximum operating life of the charging station, $C_j$ is the total construction cost of the $j$-th charging station, $C_{cp,j}$ is the purchase cost of the charging pile of the $j$-th charging station, and $C_{om,j}$ is the annual operation and maintenance cost of the $j$-th charging station.

The purchase cost of charging piles is directly related to the number and type of charging piles in the charging station. At the same time, different charging pile power configuration schemes will indirectly affect the charging waiting time of users in the charging station. Taking into account the importance of user satisfaction [33], this paper comprehensively considers the combination of fast and slow charging piles to maximize the charging satisfaction of users. The purchase cost of charging piles is as follows:

$$C_{cp,j} = \left( C_{fast} N_{fast,j} + C_{slow} N_{slow,j} \right) \tag{4}$$

where $C_{fast}$ is the purchase unit price of the fast charging pile, $N_{fast,j}$ is the configuration quantity of fast charging piles in the $j$-th charging station, $C_{slow}$ is the purchase unit price of the slow charging pile, and $N_{slow,j}$ is the configuration number of slow charging piles in the $j$-th charging station.

The operation and maintenance cost of the charging station mainly includes the maintenance cost of equipment in the charging station, the operation loss cost of equipment, and the salary of staff [34]. Usually, the specific value of each cost is vague and difficult to calculate. Therefore, in this paper, the annual operation and maintenance cost of the charging station is converted according to the percentage of the initial investment of the charging station. If the conversion factor is $\mu$, the annual operation and maintenance cost of the $j$-th charging station can be expressed by the following formula:

$$C_{om,j} = \mu \left( C_j + C_{cp,j} \right) = \mu \left[ C_j + \left( C_{fast} N_{fast,j} + C_{slow} N_{slow,j} \right) \right] \tag{5}$$

**2.1.2 The user's charging additional cost**

The user's charging additional cost mainly includes the time cost of driving to the target charging station, the queuing time loss cost in the charging station, and the power consumption cost of driving on the road. Therefore, the calculation formula is as follows:

$$\min f_2 = C_{time1} + C_{time2} + C_{loss} \tag{6}$$

where $C_{time1}$ is the equivalent economic cost of the time spent by the user when driving to the charging station, $C_{time2}$ is the equivalent economic cost of the waiting time of users in the charging station, $C_{loss}$ is the equivalent economic cost of the electricity consumed by the user when driving to the charging station.

The time cost of driving to the target charging station is closely related to the charging path selection and the driving speed of the electric vehicle, so the calculation formula is as follows:

$$C_{time1} = T_{year} \cdot k \cdot \sum_{j=1}^{N} \sum_{i=1}^{D} \frac{\alpha NUM_i d_{ij}}{v_i} \tag{7}$$

where $k$ is the time value of user travel, which can be calculated from the average income of residents in the planning area. $v_i$ is the driving speed of the user when passing the $i$-th road. $d_{ij}$ is the average driving distance from the $i$-th road to the $j$-th charging station.

The calculation of queuing time in the station is directly related to the arrival of electric vehicles in the coverage area of the charging station. Therefore, the calculation formula of queuing time cost in the station is as follows:

$$C_{time2} = T_{year} \cdot k \cdot \sum_{j=1}^{N} \sum_{t=1}^{24} D_{tj} \Delta T_{tj} \tag{8}$$

where $D_{tj}$ is the total number of electric vehicles arriving at the $j$-th charging station at time $t$. $\Delta T_{tj}$ is the average charging queuing time of electric vehicles in the $j$-th charging station at time $t$.

The equivalent economic cost of the electricity consumed by the user when driving to the charging station can be expressed by the following formula:

$$C_{loss} = T_{year} \cdot c_s \cdot g \cdot \sum_{j=1}^{N} \sum_{i=1}^{D} \alpha NUM_i d_{ij} \tag{9}$$

where $g$ is the amount of electricity consumed per unit mileage by the electric vehicle.

To sum up, this paper considers multiple objectives when planning, which can balance the operator's benefit and the user's charging additional cost and can reflect the actual problem more objectively. At the same time, in order to match with the following algorithm, this section rewrites the objective function model, the specific form is as follows:

$$\begin{cases} \min f_1 = C_1 - I \\ \min f_2 = C_{time1} + C_{time2} + C_{loss} \end{cases} \tag{10}$$

**2.2 Constraint**

This paper comprehensively considers the constraints of the rated capacity of the charging station, the peak charging power of the charging station, the number of charging piles in the charging station, the number of charging stations, and the distance between charging stations. The specific descriptions are as follows.

**2.2.1 Total rated capacity constraints of charging stations**

To meet the charge needs of all users in the planning area, the rated total capacity of charging stations should at least be greater than or equal to the maximum charging demand of electric vehicles.

$$T_d \cdot \sum_{j=1}^{N} S_j \geq \sum_{i=1}^{D} \alpha NUM_i cap \left( SOC_{ref} - SOC_0 \right) \tag{11}$$

$$S_j = P_{fast} N_{fast,j} + P_{slow} N_{slow,j} \tag{12}$$

where $T_d$ is the daily average running time of the charging piles in the charging station, $S_j$ is the rated capacity of the $j$-th charging station, $P_{fast}$ is the rated power of the fast charging pile, and $P_{slow}$ is the rated power of the slow charging pile.

**2.2.2 Peak charging power constraint of charging stations**

In order to ensure the safe operation of each charging station during the peak charging period, the rated power of the $j$-th charging station shall be greater than or equal to the peak charging power of electric vehicles.

$$S_j \geq \sum_{i} \alpha \beta \eta NUM_{ij} cap \left( SOC_{ref} - SOC_0 \right) \tag{13}$$

where $\beta$ is the simultaneous charging rate of electric vehicles, $\eta$ is the multiple relationship between the peak and average charging power of electric vehicles in the planning area, and $NUM_{ij}$ is the total number of electric vehicles on the $i$-th road covered by the $j$-th charging station.

**2.2.3 Restriction on the number of charging piles in the station**

To ensure the service quality of the charging station and reduce the resources waste caused by the number accumulation of charging piles in the station, this paper restricts the configuration quantity of

the fast and slow charging piles in the station. The specific formulas are as follows:

$$N_{fast,\min} \leq N_{fast,j} \leq N_{fast,\max} \tag{14}$$

$$N_{slow,\min} \leq N_{slow,j} \leq N_{slow,\max} \tag{15}$$

where $N_{fast,\max}$ and $N_{fast,\min}$ are the upper and lower limits of the number of fast charging piles configured in the charging station. This article refers to the relevant content of reference [35], and the values are set to 35 and 10, respectively. The upper and lower limits of the number of slow charging piles are the same as fast charging piles. $N_{slow,\max}$ and $N_{slow,\min}$ are the upper and lower limits of the number of slow charging piles configured in a charging station.

**2.2.4 Constraints on the number of charging stations**

On the one hand, the number of charging stations directly affects the operator's income. On the other hand, the user's charging satisfaction level is also directly related to the number of charging stations, so it is necessary to reasonably plan the number of charging stations constructed in the planning area. The number of charging stations mainly depends on the total charging demand in the planning area and the upper and lower limits of the capacity of the charging station. The following constraints are established:

$$N_{\min} \leq N \leq N_{\max} \tag{16}$$

$$N_{\max} = ceil\left(\frac{\sum_{i=1}^{D} \alpha NUM_i cap\left(SOC_{ref} - SOC_0\right)}{T_d N_{slow,\min} P_{slow}}\right) \tag{17}$$

$$N_{\min} = ceil\left(\frac{\sum_{i=1}^{D} \alpha NUM_i cap\left(SOC_{ref} - SOC_0\right)}{T_d N_{fast,\max} P_{fast}}\right) \tag{18}$$

where $N_{\min}$ is the minimum number of charging stations built in the planning area, $N_{\max}$ is the maximum number of charging stations built in the planning area, and $ceil$ () is a round-up function.

**2.2.5 Constraints on the construction distance of charging stations**

In order to meet the daily travel needs of users and ensure the driving ability of electric vehicles, the construction distance between adjacent charging stations should not be too far. At the same time, in order to avoid the waste of charging resources, the actual distance between adjacent charging stations should not be too close [36]. Therefore, the service radius of charging stations and the actual distance of adjacent charging stations should obey the following constraints:

$$r_j \leq S_{j,j+1} \leq 2r_j \tag{19}$$

where $r_j$ is the effective service radius of the $j$-th charging station, and $S_{j,j+1}$ is the actual distance between the $j$-th charging station and the $j+1$-th charging station.

**2.2.6 Restriction on the number of electric vehicles served by the charging station**

In order to effectively ensure the charging comfort of users, the number of electric vehicles served by the charging station shall be within a certain limit [37], so the following constraints are set:

$$B_{\min} \leq \delta_j \leq B_{\max} \tag{20}$$

$$\delta_j = \sum_i NUM_{ij} \tag{21}$$

where $B_{\max}$ and $B_{\min}$ are the upper and lower limits of the number of electric vehicles served by the charging station respectively, and this paper calculates the value of the two based on reference [12]. $\delta_j$ is the total number of electric vehicles served by the *j*-th charging station.

**2.2.7 Distance constraint between the charging demand point and the charging station**

In order to ensure the service quality of the charging station and maximize the charging satisfaction of users, the distance between the *i*-th charging demand point and the *j*-th charging station should be less than or equal to the effective service radius of the *j*-th charging station [35].

$$d_{ij} \leq r_j \tag{22}$$

**2.2.8 The charging characteristic constraints of electric vehicles**

In the rapid charging process of electric vehicles, the charging power usually starts with high power and then decreases as the battery SOC approaches 100%, resulting in a convex function relationship between the charging power of electric vehicles and time. According to the charging characteristics of electric vehicles, in order to ensure the charging efficiency, reduce the queuing time of users and ensure the service life of electric vehicles, this paper restricts the initial value and the threshold of SOC during charging with reference [38]. The formulas are as follows:

$$SOC_0 \geq 20\% \tag{23}$$

$$SOC_{ref} \leq 90\% \tag{24}$$

**2.2.9 Maximum allowable charging power constraint of the distribution network**

Considering the safe operation of the distribution network in the planning area, the total rated power of all charging stations shall be less than or equal to the maximum charging power allowed by the distribution network.

$$\sum_{j=1}^{N}\left(P_{fast}N_{fast,j} + P_{slow}N_{slow,j}\right) \leq P_{max} \tag{25}$$

where $P_{max}$ is the maximum charging power allowed to be connected to the distribution network.

**2.3 The influence between road traffic conditions and the user's charging additional cost**

It can be seen from the above model that the road loss time cost generated during the user's charging process has a great relationship with the driving speed of the electric vehicle. However, when designing the planning scheme of charging stations, previous studies only selected the average driving speed of electric vehicles to analyze the charging behavior and charging satisfaction of users, which is not universal. At the same time, this will produce a large deviation between the results and the actual situation, resulting in the poor economy of the obtained scheme. Therefore, how to reasonably analyze the driving speed of electric vehicles will help reduce the user's charging additional cost and further improve the charging satisfaction of users.

We can know that the performance of car energy consumption is closely related to road congestion. Usually, the road is more congested, and the fuel consumption of the car is higher. For electric vehicles, the road traffic is more congested, the user travel is slower, and the travel time is longer. Combined with formula (7), it can be seen that the cost of road loss time incurred by the user during the charging process is greater, resulting in poorer charging satisfaction of the user. Taking Beijing as an example, the corresponding relationship between road congestion and user travel time is shown in the following table [39]:

Table 1  The relationship between road congestion and user travel time in Beijing

| Road congestion level | Road congestion | User drive time |
|---|---|---|
| 0 | clear | It can be measured according to the normal speed |
| 1 | basically unblocked | It is 1.3 to 1.5 times the driving time when the road is clear |
| 2 | slightly unblocked | It is 1.5 to 1.8 times the driving time when the road is clear |

| 3 | moderate congestion | It is 1.8~2.0 times the driving time when the road is clear |
| 4 | serious congestion | It is 2 times or more of the driving time when the road is clear |

It can be seen from the above table that the road congestion level is higher, the traffic congestion is more serious, and the travel time in this traffic condition is longer. In this paper, the road time-consuming index is defined as $\lambda_i$, which means that the actual time required for the user to pass the road is a multiple of the time spent when the road is clear. For example, when $\lambda_i=1.8$, it means that the user's driving time is 1.8 times as long as the time spent in clear traffic conditions. Assuming that the driving speed of the electric vehicle is $v_0$ when the road is clear, the driving speed of the user on the road can be expressed by

$$v_i = \frac{v_0}{\lambda_i} \tag{26}$$

Formulas (7) and (26) are combined to quantify the impact of road traffic conditions on the time cost spent by users when driving to the target charging station. The calculation process is shown in formula (27).

$$C_{time} = T_{year} \cdot k \cdot \frac{\sum_{j=1}^{N}\sum_{i=1}^{D} \lambda_i \alpha NUM_i d_{ij}}{v_0} \tag{27}$$

According to formula (27), it can be seen that the road loss time cost generated by the user during the charging process is proportional to the time-consuming index of the road. Therefore, when the road time-consuming index is larger, the driving speed of the electric vehicle will be slower, the road loss time cost generated by the user will be higher, and the additional charging cost of the user will be greater. To sum up, after considering the road congestion in the planning area, the user's driving speed through each road will no longer be a constant, but is closely related to the time-consuming index of the road. This paper analyzes and calculates the driving speed of electric vehicles on each road in the planning area, and introduces the road time-consuming index to comprehensively reflect the impact of road traffic conditions on the user's charging additional cost. This can provide theoretical support for the location and capacity of charging stations, and also conform to the charging path selection of users.

## 3 Model solving

The location and capacity of the charging station is a complex nonlinear programming problem, which is usually solved by intelligent algorithms such as genetic algorithm, particle swarm optimization algorithm, and NSGA-II algorithm. Among them, NSGA-II algorithm has been widely concerned by experts and scholars in the field of charging station planning because of its strong robustness and Pareto solution optimization ability. As one of the most classical, superior, and popular algorithms, NSGA-II algorithm has been widely used in solving multi-objective optimization problems, especially the double objective function optimization problems [40-42]. However, with the continuous expansion of the population size, the algorithm also exhibits some defects such as slow calculation speed, low convergence accuracy, and poor distribution uniformity of the solution set. In view of the above problems, this paper designs an improved chaotic initialization strategy and an improved crossover operator to optimize the algorithm, which will be introduced in detail below.

### 3.1 Chaos initialization strategy

The space traversal degree of the individuals of the population is not considered in the initialization of the NSGA-II algorithm, which causes the problem of poor global convergence of the algorithm. At present, in order to improve the convergence accuracy of the algorithm, experts and scholars at home and abroad usually use the search operator method or add a search strategy in a small area to make up for the poor convergence accuracy of the algorithm. On this basis, some research results have been achieved. Reference [43] improves the NSGA-II algorithm by using the mountain climbing strategy, so as to improve the searchability of the algorithm for the optimal solution in the spatial domain and increase the ergodicity of the population by expanding the local range. Through the above improvement of the algorithm, the convergence accuracy and calculation speed of the algorithm are improved in a certain sense. However, in the face of a large-scale population, the algorithm is more likely to cause an increase in similar solutions, and the reliability and effectiveness of the algorithm are low.

The chaotic phenomenon appears to be chaotic, but it actually has certain internal laws. The obvious feature of this phenomenon is its high sensitivity to the initial value, and its motion characteristics are uncertainty and space ergodicity. According to this feature, the global convergence of the algorithm in the solution process and the search ability of the population can be enhanced [44,45]. Therefore, this paper uses the logistic mapping method to optimize the initial value of each particle in the population to make it as evenly distributed in the space as possible, which is conducive to improving the global convergence ability of the algorithm. The logistic equation is a typical chaotic system, and its mapping method is shown the following equation:

$$x_{k+1} = \tau x_k (1 - x_k) \tag{28}$$

where $\tau$ is a chaotic control parameter, which generally takes a value between 0 and 4, and its value determines the chaotic degree of the logistic equation [46]. The value of $\tau$ is larger, the chaotic degree of the system is higher. $x_k$ is the chaotic variable before the update. $x_{k+1}$ is the chaotic variable after the update, both of which take values between 0 and 1.

It can be seen from the above formula that the value of $x_{k+1}$ is jointly determined by the $\tau$ and $x_k$. When the initial value changes very little, the entire sequence will change significantly. The parameter value is larger, the chaotic phenomenon is more obvious, the range of population individuals is larger, and the global convergence of the algorithm is better.

**3.2 Improvement of crossover operator**

Crossover is the most important operation step in the genetic algorithm. During the crossover process, the genes of excellent individuals can multiply and spread rapidly, which can effectively make the remaining individuals in the population move toward the optimal solution. The simulated binary crossover (SBX) is used in the evolutionary crossover process of the traditional NSGA-II algorithm. The crossover principle is to randomly select the chromosome gene intersection so that the chromosome genes on both sides of the intersection can exchange with each other. However, this crossover method can easily lead to premature convergence of the algorithm, resulting in poor population diversity and uniformity of solution set distribution. Aiming at this problem, this paper introduces the arithmetic crossover operator into its mathematical model. Compared with the SBX operator, the arithmetic crossover operator has better global search ability and can better maintain the diversity of the population. Assuming that $X_A^t$ and $X_B^t$ are the real-valued codes of the corresponding decision variables at the intersection of the two individuals in the *t*-th generation, the calculation of the decision variables corresponding to the two individuals after the crossover is shown in formula (29):

$$\begin{cases} X_A^{t+1} = mX_A^t + nX_B^t \\ X_B^{t+1} = nX_A^t + mX_B^t \end{cases} \tag{29}$$

Among

$$m + n = 1 \quad m, n \in [0,1] \tag{30}$$

where $X_A^{t+1}$ and $X_B^{t+2}$ are the real value codes of the decision variables corresponding to the two individuals after the crossover. *m*, *n* are random numbers uniformly distributed between 0 and 1.

**3.3 Model solving steps**

After the NSGA-II algorithm is optimized by the above method, it is applied to the solution process of the location and capacity model of charging stations established in this paper. The detailed steps of the solution are as follows:

1) Step 1: set initialization parameters and chromosome coding rules. Set the relevant parameters of the algorithm, including the number of iterations, the population size, the number of decision variables, the number of objective functions, the crossover rate, and the mutation rate. The goal of optimization in this paper is as shown in formula (10). The decision variables are the construction location of the charging station and the number of fast and slow charging piles installed inside the charging station. The integer coding form is set and the coding method is as follows:

$$x = \left[ \underbrace{y_1, y_2, ...y_i}_{D}, \underbrace{NF_1, NF_2, ...NF_i}_{D}, \underbrace{NS_1, NS_2, ...NS_i}_{D} \right] \quad (31)$$

where $x$ represents a chromosome. $y_i$ is a 0/1 variable, which indicates the construction of charging station at the $i$-th candidate point. When the value of $y_i$ is 1, it means that a charging station is built at the $i$-th candidate point. When the value of $y_i$ is 0, it means that no charging station is built at the $i$-th candidate point. $NF_i$ and $NS_i$ are the numbers of fast and slow charging piles installed at the $i$-th candidate point respectively, and their values are closely related to the value of $y_i$. When a charging station is built at the $i$-th candidate point, $NF_i$ and $NS_i$ can take random values within the constraints. Otherwise, the values of $NF_i$ and $NS_i$ are both 0.

2) Step 2: initialize the population. Logistic mapping is used to initialize the decision variables of each chromosome in the population so that it can traverse all the solutions in the space as much as possible.

3) Step 3: non-dominated sorting of the population. The first-generation parent population is generated by iteration, and the objective function values corresponding to all individuals in the population are calculated according to the established model. Then this paper performs Pareto sorting and crowding degree calculation on all individuals in the population, and assigns the Pareto sorting level and crowding degree calculation value of each chromosome to the new gene segment of the chromosome. The updated chromosome is shown in formula (32):

$$x = \left[ \underbrace{y_1, y_2, ...y_i}_{D}, \underbrace{NF_1, NF_2, ...NF_i}_{D}, \underbrace{NS_1, NS_2, ...NS_i}_{D}, par, dis \right] \quad (32)$$

where $par$ is the Pareto ranking of chromosome $x$, and $dis$ is the calculated crowding degree value of chromosome $x$ under this Pareto scale.

4) Step 4: tournament competition, select, crossover, and mutate the optimized population at the same time. The individuals of the population sorted in step 3 randomly participate in the tournament competition in pairs, and the better individuals are selected. According to the improved crossover operator in this paper, the optimized population is crossed and mutated to generate the first generation of offspring individuals.

5) Step 5: combine the optimized parent individuals and the generated offspring individuals to form a new parent population, perform non-dominated sorting and crowding degree calculation on them again, and use the elite retention strategy to optimize the new population. The optimization rule is to put the non-dominated Front$_1$, Front$_2$, ..., Front$_m$ into the new parent population from high to low according to the Pareto level, until the size of the parent population exceeds the specified value. Then, the individuals in the non-dominated layer are removed according to the crowding degree from low to high, until the size of the population meets the requirements.

6) Step 6: judge whether the number of iterations of the algorithm reaches the threshold. If not, repeat steps (4) ~ (6). If it meets the requirements, directly output the Pareto optimal solution set.

7) Step 7: based on the Pareto optimal solution set obtained in step 6, the entropy weight method is used to first perform data normalization on all solutions in the set [47]. Since the indicators selected in this paper are negative indexes, the standardized processing formula is as follows:

$$Y_{ij} = \frac{\max(X_j) - X_{ij}}{\max(X_j) - \min(X_j)} \quad i = 1, 2, \cdots n, j = 1, 2, \cdots k \tag{33}$$

where $X_{ij}$ is the $j$-th index of the $i$-th feasible solution in the Pareto optimal solution set. $Y_{ij}$ is the standardized data of $X_{ij}$. $\min(X_j)$ is the minimum value of the $j$-th index among all feasible solutions. $\max(X_j)$ is the maximum value of the $j$-th index among all feasible solutions.

8) Step 8: based on the results obtained in step 7, calculate the proportion of the index value of the $i$-th feasible solution under the $j$-th index. The calculation formula is as follows:

$$w_{ij} = \frac{Y_{ij}}{\sum_{i=1}^{n} Y_{ij}} \tag{34}$$

where $w_{ij}$ is the proportion of the index value of the $i$-th feasible solution under the $j$-th index.

9) Step 9: based on the results obtained in step 8, the information entropy of the $j$-th index is calculated, and the calculation formula is as follows:

$$e_j = -\frac{\sum_{i=1}^{n} w_{ij} \ln w_{ij}}{\ln n} \tag{35}$$

If the value of $w_{ij}$ is 0, the following formula is defined:

$$w_{ij} \ln w_{ij} = 0 \tag{36}$$

where $e_j$ is the information entropy of the $j$-th index.

10) Step 10: calculate the weight of the $j$-th index, the specific formula is as follows:

$$W_j = \frac{1 - e_j}{k - \sum_{j=1}^{k} e_j} \tag{37}$$

where $W_j$ is the weight of the $j$-th index.

11) Step 11: calculate the comprehensive score of each feasible solution, and obtain the optimal solution according to the score. The formula for calculating the comprehensive score is as follows:

$$S_i = \sum_{j=1}^{k} W_j Y_{ij} \tag{38}$$

where $S_i$ is the comprehensive score of the $i$-th feasible solution in the Pareto optimal solution set.

The solution process of the location and capacity planning model of the charging station is shown in Fig. 1.

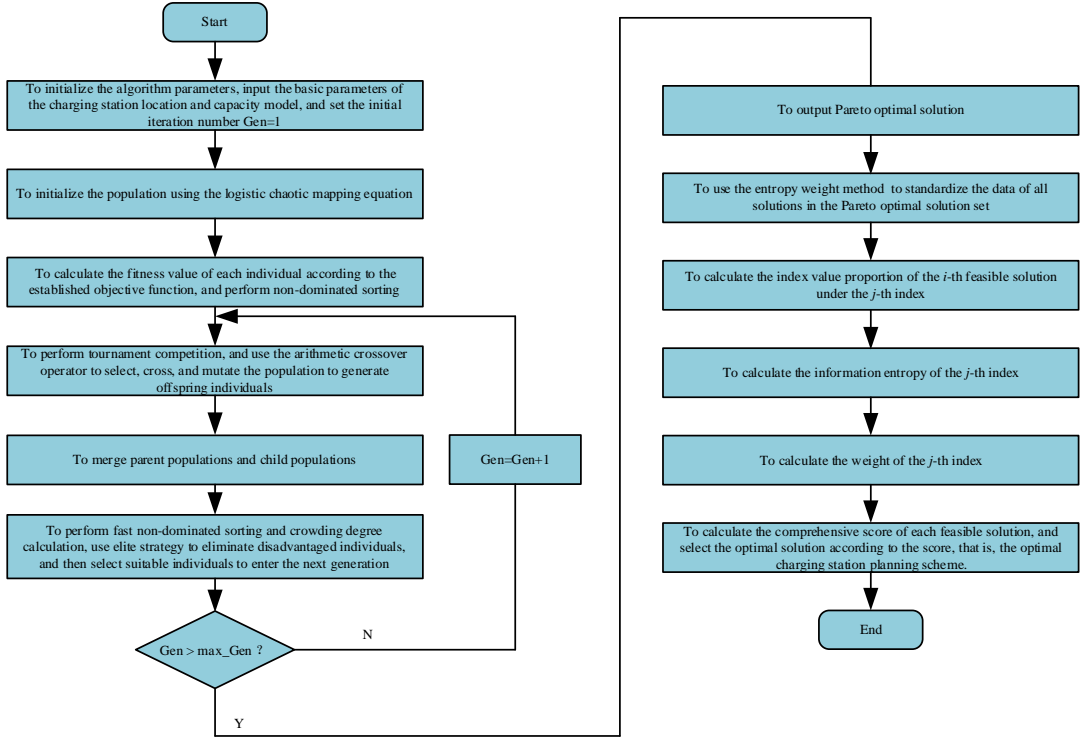

Fig.1 The flow chart for solving the location and capacity model of the charging station based on the improved NSGA-II algorithm and entropy weight method

## 4 Case and result analysis

To examine the effectiveness of the proposed approach, extensive simulation tests have been carried out in this section. The simulation platform is as follows: the processor is i5-8250U, the running memory is 8GB, and the simulation software is MATLAB R2021a.

### 4.1 Basic data

In this paper, the simulation case in reference [48] is used as a reference, and a certain area in Beijing is taken as an example for the simulation verification. Note that, in addition to using user data to simulate user behavior, charging scenarios can also be generated by using generative adversarial networks [49, 50], which can learn the distribution of electric vehicle charging sessions. The topographic map is shown in Fig. A1 of Appendix A. The area covers an area of about 100 km$^2$ and contains a total of 234 major roads. The settings of relevant parameter in the location and capacity model of the charging station established in this paper are shown in Table 2.

Table 2 The relevant parameters table of the location and capacity model of the charging station

| Parameter | Value | Parameter | Value |
| --- | --- | --- | --- |
| $D$ | 234 | $T_{year}$ | 365 |
| $c_s$ | 1yuan/(kW·h) | $c_p$ | 0.8yuan/(kW·h) |
| $P_{slow}$ | 40kW | $r$ | 0.03 |
| $C_{fast}$ | 25,000yuan | $C_{slow}$ | 10,000yuan |
| $\alpha$ | 0.08 | $g$ | 0.3 |
| $\mu$ | 0.2 | $SOC_{ref}$ | 0.9 |
| $T_d$ | 16h/d | $T$ | 5year |
| $P_{fast}$ | 80kW | $k$ | 17yuan/h |
| $cap$ | 80kWh | $SOC_0$ | 0.3 |

It is estimated that by the planning year, the number of electric vehicles in Beijing will reach 800,450, which can be converted into the traffic flow of each road in the planning area. The details are shown in

Table A1 of Appendix A. At the same time, the traffic flow of each road in the planning area is divided into five grades, as shown in Table 3. In order to more intuitively show the significant differences between different road traffic flows, this paper sets up five colors of filling in cyan, green, yellow, red, and blue according to the order of road congestion levels from small to large to show the congestion degree of each road in the planning area, as shown in Fig. 2.

Table 3   The classification of road traffic flow levels in the planned area

| Road congestion level | Road congestion | Traffic flow (vehicles/day) | Speed (km/h) |
|---|---|---|---|
| 0 | smooth | 30-100 | 40 |
| 1 | basically unblocked | 100-200 | 28 |
| 2 | light congestion | 200-300 | 25 |
| 3 | moderate congestion | 300-400 | 22 |
| 4 | serious congestion | 400-520 | 20 |

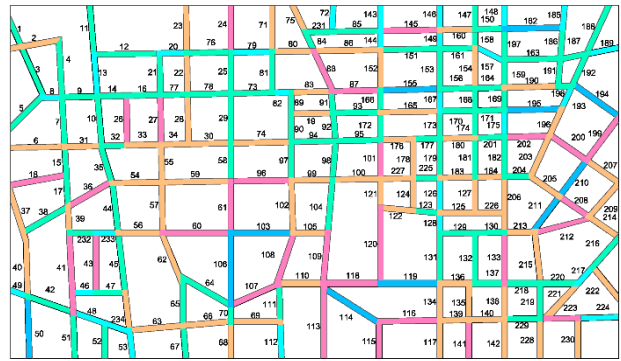

Fig.2   Comparison diagram of electric vehicle congestion of each road in the planning area

The settings of the main control parameters of the algorithm are shown in Table 4.

Table 4   The setting of main control parameters of the algorithm

| Parameter | Value |
|---|---|
| population size | 200 |
| number of iterations | 500 |
| total number of decision variables | 702 |
| number of objective functions | 2 |
| crossover probability | 0.9 |
| mutation probability | 0.1 |
| chaos control parameter | 4 |

In this paper, the population size is set to 200, the iterations number of the population is set to 500, the number of decision variables is set to 702, the number of objective functions is set to 2, the crossover probability is set to 0.9, the mutation probability is set to 0.1, and the value of the chaos control parameter is set to 4.

**4.2 Result analysis**
**4.2.1 Superiority verification of the improved NSGA-II algorithm**

To verify the advantages of the proposed algorithm, this paper uses the MPSO algorithm, the MOALO algorithm, and improved NSGA-II algorithm to solve the location and capacity model of the charging station. The learning factors of the MPSO algorithm are set to 2, and the upper and lower limits of weights are 0.9 and 0.4, respectively. At the same time, this paper sets the number of ant-lions in the MOALO algorithm to 100, the number of ants to 100, and the maximum number of iterations to 500. The Pareto optimal solution curves obtained by three algorithms are shown in Fig. 3.

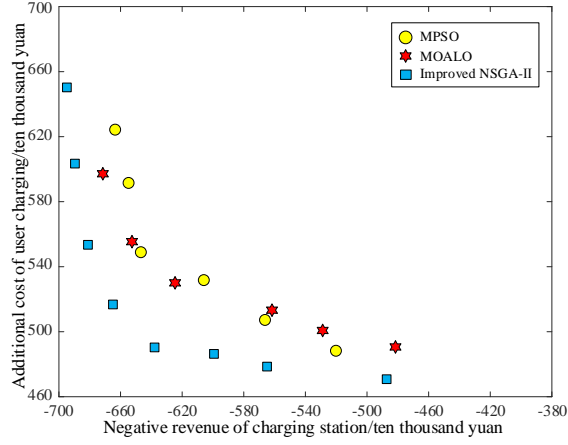

Fig.3 The Pareto optimal solution curves obtained by three algorithms

As can be seen from Fig. 3, when the income of the charging station is the same, that is, the construction scale of the charging station is the same, the additional charging cost of the user in the scheme obtained by the improved NSGA-II algorithm is the lowest, while the additional charging cost of the user in the scheme obtained by the MOALO algorithm and the MPSO algorithm is not much different, which is higher than the result obtained by the improved NSGA-II algorithm. In addition, from the distribution of the Pareto solution set, the distribution of the Pareto optimal solution curve obtained by the MOALO algorithm is concentrated in some places and sparse in some places, and the smoothness of the curve is relatively general. The Pareto optimal solution curve obtained by the MPSO algorithm and improved NSGA-II algorithm is more uniform, that is, the global convergence performance of the algorithm is better.

To further illustrate the superior performance of the algorithm proposed in this paper in terms of global convergence and computational efficiency, this paper introduces the spatial evaluation index and the total iteration time to comprehensively evaluate the algorithm. The calculation formula of the spatial evaluation index is as follows [51, 52]:

$$SP=\sqrt{\frac{1}{n-1}\sum_{i=1}^{n}(d_{av}-d_i)^2} \qquad (39)$$

Among

$$d_i = \min\left(\left|f_1^j(x)-f_1^i(x)\right|+\left|f_2^j(x)-f_2^i(x)\right|\right) \qquad i,j=1...n \qquad (40)$$

where $n$ is the total number of individuals in the solution set, and $d_{av}$ is the average value of all $d_i$ in the solution set. The definition of $d_i$ is similar to the crowding distance, and its value is the sum of the sides of the space occupied by the solution $i$ after normalization. $SP$ represents the distribution uniformity of the obtained Pareto solution set. The value is smaller, the distribution of the obtained solution is more uniform, that is, the global convergence performance of the algorithm is better.

Based on the above definition of performance evaluation index, the MOALO algorithm, MPSO algorithm, and improved NSGA-II algorithm are used to solve the model proposed in this paper independently for 10 times respectively. The comparison of $SP$ value and convergence time is shown in Table 5.

Table 5 Comparison of the performance indicators of the three algorithms in solving the model in this paper

| Serial number | SP | | | T(S) | | |
| --- | --- | --- | --- | --- | --- | --- |
| | MOALO | MPSO | Improved NSGA-II | MOALO | MPSO | Improved NSGA-II |
| 1 | 0.961 | 0.912 | 0.913 | 62.165 | 75.232 | 47.228 |
| 2 | 0.976 | 0.930 | 0.925 | 63.283 | 80.167 | 44.262 |
| 3 | 0.996 | 0.893 | 0.884 | 58.382 | 79.225 | 46.437 |
| 4 | 0.972 | 0.922 | 0.918 | 59.196 | 78.118 | 45.381 |
| 5 | 1.080 | 0.979 | 0.982 | 63.123 | 82.073 | 49.619 |
| 6 | 0.933 | 0.881 | 0.885 | 66.222 | 84.409 | 50.015 |
| 7 | 0.986 | 0.925 | 0.918 | 61.452 | 77.012 | 42.109 |
| 8 | 0.955 | 0.899 | 0.901 | 63.308 | 79.525 | 46.094 |
| 9 | 1.012 | 0.873 | 0.881 | 57.237 | 76.108 | 44.122 |
| 10 | 0.968 | 0.910 | 0.902 | 59.129 | 79.953 | 48.352 |

From the Table 5, the 10 times average value of the distribution index *SP* of the Pareto optimal solution set obtained by the MOALO algorithm is 0.984, the 10 times average value of the distribution index *SP* of the Pareto optimal solution set obtained by the MPSO algorithm is 0.912, and the 10 times average value of the *SP* obtained by the improved NSGA-II algorithm is 0.911. The *SP* values of the improved NSGA-II algorithm and the MPSO algorithm are similar, which are significantly smaller than the results obtained by the MOALO algorithm. Therefore, the distribution uniformity of the solution sets of the two algorithms is better than that of the MOALO algorithm. In addition, compared with MOALO algorithm and MPSO algorithm, the improved NSGA-II algorithm requires a shorter iteration time in solving the model in this paper. Although the MPSO algorithm has good global convergence ability, it requires a long iteration time, that is, the computational efficiency is not high. In conclusion, compared with the MOALO algorithm and MPSO algorithm, the improved NSGA-II algorithm proposed in this paper has better performance.

**4.2.2 Analysis of the impact of road traffic conditions on the planning scheme**

In order to analyze the impact of the road traffic conditions on the charging station planning scheme, this part sets up two cases with and without considering the road traffic conditions. The Pareto optimal solution curves of the two cases are shown in Fig. 4.

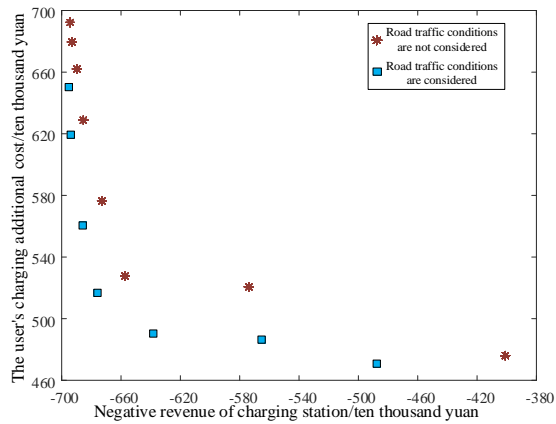

Fig.4 The Pareto optimal solution curves of two cases

Using the entropy weight method, the obtained Pareto optimal solution curves of two cases are objectively evaluated and scored, and the scoring results are shown in Table 6 and Table 7.

Table 6 The scoring of all solutions under road traffic conditions is considered

| Number | Negative value of charging station revenue /yuan | The user's charging additional cost /yuan | Score |
|---|---|---|---|
| 1 | -6,950,300 | 6,505,000 | 0.436 |
| 2 | -6,937,100 | 6,189,700 | 0.532 |
| 3 | -6,873,800 | 5,567,700 | 0.713 |
| 4 | -6,768,700 | 5,164,600 | 0.817 |
| 5 | -6,387,000 | 4,876,600 | 0.819 |
| 6 | -5,659,700 | 4,852,200 | 0.682 |

| 7 | -4,877,200 | 4,702,200 | 0.564 |

Table 7  The scoring of all solutions under road traffic conditions is not considered

| Number | Negative value of charging station revenue /yuan | The user's charging additional cost /yuan | Score |
|---|---|---|---|
| 1 | -6,944,900 | 6,921,100 | 0.279 |
| 2 | -6,935,900 | 6,798,500 | 0.319 |
| 3 | -6,899,400 | 6,614,500 | 0.377 |
| 4 | -6,854,000 | 6,285,900 | 0.482 |
| 5 | -6,735,500 | 5,659,500 | 0.678 |
| 6 | -6,577,900 | 5,275,400 | 0.794 |
| 7 | -5,741,900 | 5,202,700 | 0.735 |
| 8 | -4,016,100 | 4,749,200 | 0.721 |

Based on the above scoring, considering the road traffic conditions, the optimal solution is the scheme corresponding to No. 5 in Table 6. Without considering the traffic conditions, the optimal solution is the scheme corresponding to No. 6 in Table 7. The cost comparisons of the two schemes are shown in Table 8 and Table 9.

Table 8  The each cost of the plan under road traffic conditions is considered

| Number | Annual construction cost of charging stations /yuan | Annual purchase cost of charging piles /yuan | Annual operation and maintenance cost of charging stations /yuan | Total annual profit of charging stations /yuan |
|---|---|---|---|---|
| 23 | 3,399,700 | 1,158,200 | 911,600 | 6,387,000 |
| User travel loss time cost /yuan | User queuing time cost at charging stations /yuan | Total user time loss cost /yuan | Cost of electricity consumed by users on the road /yuan | The user's charging additional cost /yuan |
| 2,109,100 | 1,314,700 | 3,423,800 | 1,452,800 | 4,876,600 |

Table 9  The each cost of the plan under road traffic conditions is not considered

| Number | Annual construction cost of charging stations /yuan | Annual purchase cost of charging piles /yuan | Annual operation and maintenance cost of charging stations /yuan | Total annual profit of charging stations /yuan |
|---|---|---|---|---|
| 22 | 3,251,900 | 1,146,900 | 879,800 | 6,577,900 |
| User travel loss time cost /yuan | User queuing time cost at charging stations /yuan | Total user time loss cost /yuan | Cost of electricity consumed by users on the road /yuan | The user's charging additional cost /yuan |
| 2,380,400 | 1,368,800 | 3,749,200 | 1,526,200 | 5,275,400 |

Through the cost comparison analysis of the two schemes, it can be seen that when the road traffic conditions are not considered, the user's road loss time cost is 2,380,400 yuan, the total additional cost during charging is 5,275,400 yuan, and the charging station operator's income is 6,577,900 yuan. After considering the road traffic conditions, the user's road loss time cost is 2,109,100 yuan, the total additional cost during charging is 4,876,600 yuan, and the charging station operator's income is 6,387,000 yuan. Compared with the scheme without considering road traffic conditions, although the income of the operator in this scheme is slightly reduced by about 2.9%, the user's road loss time cost is significantly reduced by 11.4% and the user's charging additional cost is also reduced by 7.6%. This is because this scheme considers the traffic flow of each road in the planning area and the driving speed of electric vehicles on each road is graded and refined, which improves the defect of setting the average driving speed in the traditional model. While ensuring the optimal economy of the solution, it also improves the user's charging satisfaction to a certain extent. At the same time, it is more suitable for the actual situation, which is conducive to the actual promotion and application of the solution. Table 10 and Table11 are the detailed charging station layouts obtained by the two schemes.

Table 10  The planning scheme of charging stations with considering road traffic conditions

| Number | Number of fast charging piles | Number of slow charging piles | Capacity of charging stations /kW | Number | Number of fast charging piles | Number of slow charging piles | Capacity of charging stations /kW |
| --- | --- | --- | --- | --- | --- | --- | --- |
| 4 | 13 | 21 | 1,880 | 116 | 19 | 11 | 1,960 |
| 18 | 15 | 18 | 1,920 | 120 | 22 | 10 | 2,160 |
| 24 | 10 | 16 | 1,440 | 125 | 13 | 20 | 1,840 |
| 27 | 15 | 12 | 1,680 | 141 | 16 | 13 | 1,800 |
| 43 | 22 | 14 | 2,320 | 145 | 12 | 16 | 1,600 |
| 50 | 16 | 12 | 1,760 | 155 | 15 | 18 | 1,920 |
| 54 | 13 | 16 | 1,680 | 162 | 16 | 20 | 2,080 |
| 88 | 16 | 20 | 2,080 | 195 | 20 | 13 | 2,120 |
| 96 | 18 | 15 | 2,040 | 199 | 23 | 20 | 2,640 |
| 101 | 14 | 18 | 1,840 | 211 | 20 | 16 | 2,240 |
| 106 | 24 | 15 | 2,520 | 223 | 15 | 21 | 2,040 |
| 113 | 23 | 16 | 2,480 | | | | |

Table 11  The planning scheme of charging stations without considering road traffic conditions

| Number | Number of fast charging piles | Number of slow charging piles | Capacity of charging stations /kW | Number | Number of fast charging piles | Number of slow charging piles | Capacity of charging stations /kW |
| --- | --- | --- | --- | --- | --- | --- | --- |
| 3 | 14 | 18 | 1,840 | 109 | 25 | 17 | 2,680 |
| 26 | 13 | 22 | 1,920 | 115 | 25 | 14 | 2,560 |
| 36 | 12 | 16 | 1,600 | 125 | 13 | 16 | 1,680 |
| 43 | 16 | 18 | 2,000 | 133 | 22 | 18 | 2,480 |
| 50 | 20 | 15 | 2,200 | 145 | 23 | 14 | 2,400 |
| 66 | 10 | 18 | 1,520 | 156 | 14 | 14 | 1,680 |
| 74 | 17 | 22 | 2,240 | 162 | 23 | 19 | 2,600 |
| 76 | 16 | 13 | 1,800 | 194 | 17 | 18 | 2,080 |
| 87 | 15 | 21 | 2,040 | 195 | 25 | 14 | 2,560 |
| 101 | 20 | 13 | 2,120 | 208 | 24 | 16 | 2,560 |
| 103 | 25 | 18 | 2,720 | 223 | 17 | 19 | 2,120 |

To effectively reduce the time cost and power loss cost of users when charging, both schemes build charging stations on the road sections with relatively congested roads, such as sections numbered 43, 50, 162, 195, etc., which can minimize the additional cost for users when charging and improve the economy of the scheme. However, compared with not considering road traffic conditions, more charging stations and larger charging capacity are built in areas with more congested roads after considering road traffic conditions, such as the areas included in the road sections numbered 96, 106, 113, 116, 120, and 141. In areas with relatively unobstructed roads, the configuration of charging stations is just the opposite. For example, in the areas included in the road sections numbered 145, 155, 162 and 195, the number and capacity of charging stations in the area are significantly reduced after considering the road traffic conditions, which makes the distribution of charging stations in the planning area more in line with the actual situation. At the same time, it can also reduce the "charging anxiety" of users to a certain extent and the users' charging additional cost.

**4.2.3 Comparative analysis of results under different charging pile configuration schemes**

To analyze the influence of the combined installation of fast and slow charging piles on the optimization results, this paper sets up the following three scenarios for comparative analysis: 1) Only slow charging piles are installed; 2) Only fast charging piles are installed; 3) Combined installation of fast and slow charging piles. The location results of charging stations under three scenarios are shown in Fig. 5. The detailed charging stations planning schemes obtained by scenario 1 and scenario 2 are shown in Table 12 and Table 13, respectively.

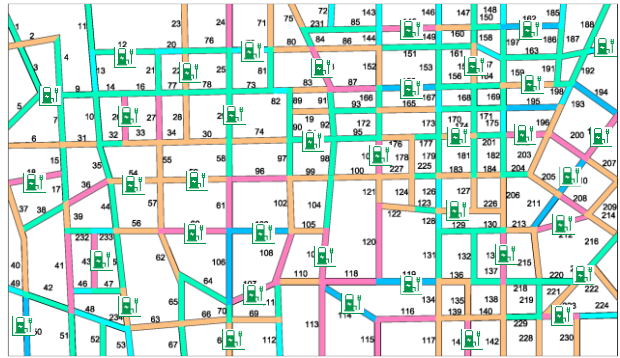

(a) The site selection results of charging stations in scenario 1

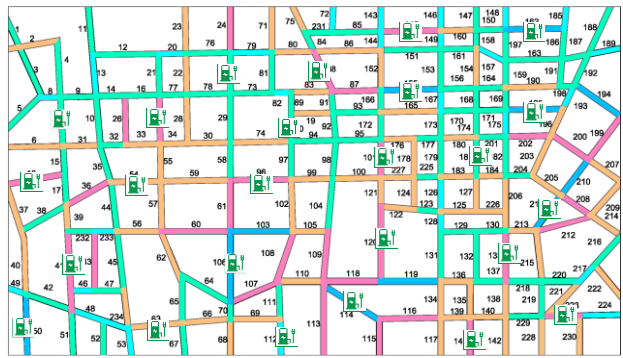

(b) The site selection results of charging stations in scenario 2

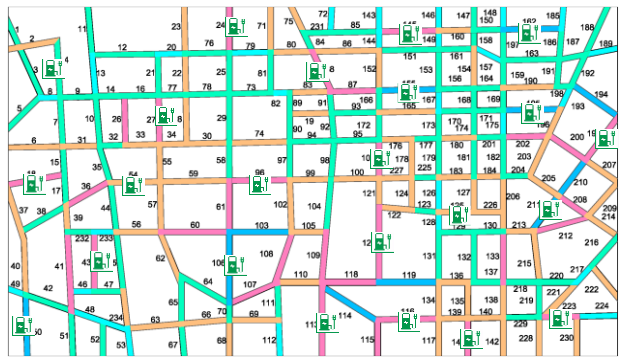

(c) The site selection results of charging stations in scenario 3

Fig.5  Display of site selection results of charging stations under the three schemes

Table 12  The planning scheme of charging stations in scenario1

| Number | Number of slow charging piles | Capacity of charging stations /kW | Number | Number of slow charging piles | Capacity of charging stations /kW | Number | Number of slow charging piles | Capacity of charging stations /kW |
|---|---|---|---|---|---|---|---|---|
| 8 | 21 | 840 | 88 | 35 | 1,400 | 155 | 33 | 1,320 |
| 12 | 25 | 1,000 | 94 | 26 | 1,040 | 162 | 34 | 1,360 |
| 18 | 30 | 1,200 | 101 | 33 | 1,320 | 180 | 31 | 1,240 |
| 22 | 32 | 1,280 | 103 | 31 | 1,240 | 189 | 21 | 840 |
| 26 | 34 | 1,360 | 107 | 33 | 1,320 | 190 | 33 | 1,320 |
| 29 | 22 | 880 | 109 | 35 | 1,400 | 199 | 30 | 1,200 |
| 43 | 32 | 1,280 | 114 | 34 | 1,360 | 202 | 33 | 1,320 |
| 50 | 34 | 1,360 | 119 | 33 | 1,320 | 210 | 32 | 1,280 |
| 54 | 29 | 1,160 | 125 | 29 | 1,160 | 212 | 32 | 1,280 |
| 59 | 28 | 1,120 | 133 | 32 | 1,280 | 217 | 28 | 1,120 |
| 60 | 34 | 1,360 | 141 | 34 | 1,360 | 223 | 32 | 1,280 |
| 68 | 29 | 1,160 | 145 | 35 | 1,400 | 234 | 28 | 1,120 |
| 79 | 30 | 1,200 | 154 | 29 | 1,160 | | | |

Table 13  The planning scheme of charging stations in scenario 2

| Number | Number of fast charging piles | Capacity of charging stations /kW | Number | Number of fast charging piles | Capacity of charging stations /kW | Number | Number of fast charging piles | Capacity of charging stations /kW |
|---|---|---|---|---|---|---|---|---|
| 7 | 14 | 1,120 | 90 | 15 | 1,200 | 145 | 25 | 2,000 |
| 18 | 22 | 1,760 | 96 | 25 | 2,000 | 155 | 28 | 2,240 |
| 25 | 17 | 1,360 | 101 | 23 | 1,840 | 162 | 29 | 2,320 |
| 27 | 25 | 2,000 | 106 | 28 | 2,240 | 181 | 18 | 1,440 |
| 41 | 27 | 2,160 | 112 | 20 | 1,600 | 195 | 26 | 2,080 |
| 50 | 21 | 1,680 | 114 | 28 | 2,240 | 211 | 30 | 2,400 |
| 54 | 19 | 1,520 | 120 | 26 | 2,080 | 223 | 28 | 2,240 |
| 63 | 20 | 1,600 | 133 | 20 | 1,600 | | | |
| 88 | 24 | 1,920 | 141 | 21 | 1,680 | | | |

It can be seen from Fig. 5(a) and Table 12 that the constructed number of charging stations is the largest when only slow charging piles are installed. At the same time, it can be seen from Fig. 5(a) that the distribution density of charging stations in scenario 1 is higher than that in the other two scenarios and the distance the user travels to the target charging station is shorter. This is conducive to reducing the user's charging additional cost and improving the user's charging comfort. However, this unilateral measure to improve the user's charging satisfaction by building a large number of charging stations will inevitably lead to a substantial increase in the construction cost of charging stations. The calculation shows that the construction cost of the charging stations in scenario 1 is 5,616,900 yuan, which is 65.21% higher than the result of scenario 3. Simultaneously, it will put great pressure on the early investment of the charging stations operator. In addition, the income of charging stations operators has been reduced from 6,387,000 million yuan in scenario 3 to 3,475,900 million yuan, which will reduce the investment enthusiasm of charging stations operators to a certain extent and delay the construction process of charging stations.

It can be seen from Fig. 5(b) and Table 13 that the constructed number of charging stations when only fast charging piles are installed ranks second. However, due to the high unit price of fast charging piles, the calculation results show that the purchase cost of charging piles in scenario 2 is 1,688,100 million yuan, which is an increase of 23.35% and 45.75% compared with scenarios 1 and 3, respectively. The total annual profit of the charging stations in the scenario 2 is 5,396,400 million yuan, which is a decrease of 15.51% compared with the scenario 3. This shows that the unilateral construction of fast charging piles in scenario 2 not only does not save the total cost of the charging stations but also significantly increases the purchase cost of the charging piles, resulting in lower economics of the planning scheme. In addition, there are blank areas for the installation of charging stations in the areas included in the road sections numbered 11, 13, 22, and 23 in the upper left corner of Fig. 5(b), which will result in uneven distribution of charging resources and seriously affect the overall charging satisfaction of users in the target area.

It can be seen from Fig. 5(c) and Table 10 that the solution obtained in scenario 3 can make a reasonable ratio of fast and slow charging piles in combination with the charging demand in each area, making the purchase cost of charging piles inside the charging station more economical. For example, in the road sections numbered 106, 113, 120, and 199 in Fig. 5(c), the area included in this road section has a relatively high charging demand, so the number of fast charging piles is much more than that of slow charging piles when building charging stations in this area, which helps to improve charging efficiency and reduce the occurrence of queues in the station. In addition, the road sections numbered 4, 24, and 125 in Fig. 5(c) contain relatively low charging demand. Therefore, the number of slow charging piles in the construction of charging stations in this area is more than that of fast charging piles, which can improve the economy of charging stations construction while ensuring user charging satisfaction. In general, the optimized solution obtained by the combined installation of fast and slow charging piles has a more uniform distribution of charging stations, a more reasonable configuration capacity of charging piles, and a better economy of the resulting solution.

## 5 Conclusions

The rational planning of the charging stations will play an irreplaceable role in the popularization and development of electric vehicles. This paper establishes a charging station location and capacity model considering the interests of both operators and users and proposes an improved NSGA-II algorithm to solve the model. The conclusions are as follows:

1) The MPSO algorithm, the MOALO algorithm and the improved NSGA-II algorithm are used to solve the model established in this paper, respectively. The results show that compared with the MPSO algorithm and the MOALO algorithm, the improved NSGA-II algorithm obtains the lowest $SP$ average of 10 times, so the solution set distribution obtained by this algorithm is more uniform, that is, the global convergence ability of the algorithm is better. In addition, in terms of model solving efficiency, compared with the MPSO algorithm and MOALO algorithm, the iteration time required by the improved NSGA-II algorithm is shortened by 41.45% and 24.43% respectively, that is, the improved NSGA-II algorithm has higher computational efficiency.

2) In the construction of the location and capacity model of charging stations, the road time-consuming index is introduced to quantify the impact of the actual road conditions on the user's charging additional cost. After considering the impact of road traffic conditions, although the income of the charging stations operator is slightly reduced by about 2.9%, the user's road loss time cost is significantly reduced by 11.4%, and the total user's charging additional cost is reduced by 7.6%. This guarantees the interests of charging stations operators and effectively improves user's charging satisfaction.

3) In terms of the charging station configuration, the combination of fast and slow charging piles is considered. Simultaneously, a reasonable ratio of fast and slow charging piles can be carried out according to the charging demand in each area, which effectively reduces the investment cost and resource waste caused by the use of a single charging pile configuration method. Therefore, the charging stations layout of the obtained optimization scheme is more uniform and the economy of the scheme is also better.

## Appendix A

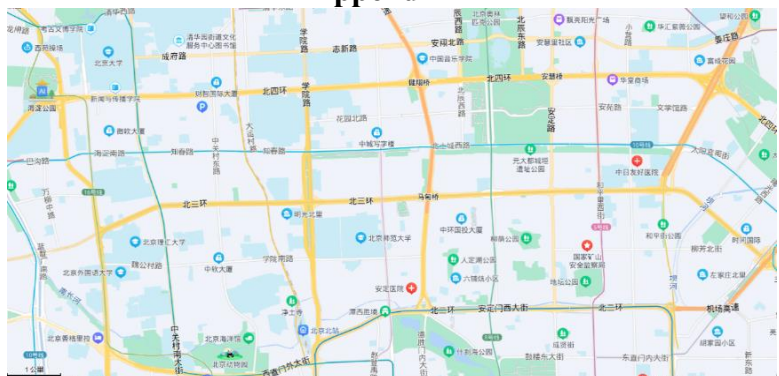

Fig.A1　The topographic map of planning area

Table A1  Statistics of electric vehicle flow on each road in the planning area

| Number | Daily traffic flow | Number | Daily traffic flow | Number | Daily traffic flow | Number | Daily traffic flow | Number | Daily traffic flow | Number | Daily traffic flow |
|---|---|---|---|---|---|---|---|---|---|---|---|
| 1 | 127 | 41 | 355 | 81 | 184 | 121 | 216 | 161 | 121 | 201 | 184 |
| 2 | 236 | 42 | 162 | 82 | 75 | 122 | 218 | 162 | 487 | 202 | 306 |
| 3 | 197 | 43 | 345 | 83 | 253 | 123 | 172 | 163 | 156 | 203 | 275 |
| 4 | 193 | 44 | 112 | 84 | 155 | 124 | 290 | 164 | 171 | 204 | 147 |
| 5 | 121 | 45 | 135 | 85 | 187 | 125 | 215 | 165 | 293 | 205 | 277 |
| 6 | 271 | 46 | 84 | 86 | 208 | 126 | 80 | 166 | 118 | 206 | 230 |
| 7 | 150 | 47 | 109 | 87 | 367 | 127 | 225 | 167 | 72 | 207 | 286 |
| 8 | 35 | 48 | 77 | 88 | 398 | 128 | 44 | 168 | 141 | 208 | 302 |
| 9 | 118 | 49 | 82 | 89 | 118 | 129 | 125 | 169 | 118 | 209 | 237 |
| 10 | 115 | 50 | 445 | 90 | 178 | 130 | 181 | 170 | 165 | 210 | 433 |
| 11 | 118 | 51 | 140 | 91 | 165 | 131 | 124 | 171 | 134 | 211 | 442 |
| 12 | 158 | 52 | 108 | 92 | 118 | 132 | 197 | 172 | 190 | 212 | 370 |
| 13 | 94 | 53 | 88 | 93 | 144 | 133 | 300 | 173 | 112 | 213 | 274 |
| 14 | 102 | 54 | 231 | 94 | 181 | 134 | 270 | 174 | 191 | 214 | 296 |
| 15 | 159 | 55 | 264 | 95 | 161 | 135 | 221 | 175 | 59 | 215 | 292 |
| 16 | 134 | 56 | 222 | 96 | 320 | 136 | 267 | 176 | 227 | 216 | 118 |
| 17 | 174 | 57 | 255 | 97 | 197 | 137 | 152 | 177 | 109 | 217 | 32 |
| 18 | 309 | 58 | 118 | 98 | 127 | 138 | 208 | 178 | 237 | 218 | 180 |
| 19 | 236 | 59 | 234 | 99 | 250 | 139 | 267 | 179 | 102 | 219 | 147 |
| 20 | 72 | 60 | 371 | 100 | 240 | 140 | 283 | 180 | 242 | 220 | 171 |
| 21 | 169 | 61 | 398 | 101 | 306 | 141 | 318 | 181 | 112 | 221 | 246 |
| 22 | 225 | 62 | 253 | 102 | 275 | 142 | 233 | 182 | 165 | 222 | 206 |
| 23 | 243 | 63 | 225 | 103 | 470 | 143 | 59 | 183 | 225 | 223 | 325 |
| 24 | 300 | 64 | 191 | 104 | 174 | 144 | 53 | 184 | 202 | 224 | 91 |
| 25 | 169 | 65 | 169 | 105 | 231 | 145 | 303 | 185 | 96 | 225 | 144 |
| 26 | 334 | 66 | 253 | 106 | 490 | 146 | 56 | 186 | 152 | 226 | 224 |
| 27 | 327 | 67 | 159 | 107 | 351 | 147 | 112 | 187 | 106 | 227 | 249 |
| 28 | 258 | 68 | 295 | 108 | 352 | 148 | 53 | 188 | 159 | 228 | 239 |
| 29 | 143 | 69 | 255 | 109 | 365 | 149 | 60 | 189 | 82 | 229 | 184 |
| 30 | 256 | 70 | 130 | 110 | 234 | 150 | 163 | 190 | 277 | 230 | 272 |
| 31 | 202 | 71 | 222 | 111 | 162 | 151 | 141 | 191 | 180 | 231 | 183 |
| 32 | 138 | 72 | 53 | 112 | 82 | 152 | 206 | 192 | 121 | 232 | 121 |
| 33 | 206 | 73 | 122 | 113 | 346 | 153 | 186 | 193 | 53 | 233 | 96 |
| 34 | 209 | 74 | 227 | 114 | 445 | 154 | 267 | 194 | 480 | 234 | 53 |
| 35 | 161 | 75 | 258 | 115 | 317 | 155 | 467 | 195 | 437 | | |
| 36 | 365 | 76 | 124 | 116 | 380 | 156 | 174 | 196 | 239 | | |
| 37 | 215 | 77 | 128 | 117 | 218 | 157 | 91 | 197 | 112 | | |
| 38 | 181 | 78 | 133 | 118 | 339 | 158 | 110 | 198 | 109 | | |
| 39 | 224 | 79 | 94 | 119 | 515 | 159 | 133 | 199 | 331 | | |
| 40 | 267 | 80 | 240 | 120 | 312 | 160 | 214 | 200 | 389 | | |